\documentstyle[psfig,prb,aps]{revtex}
\begin{document}
\draft
\title{Variational quantum Monte Carlo calculations for solid surfaces}
\author{R. Bahnsen\cite{correspond}, H. Eckstein, and W. Schattke}
\address{Institut f\"ur Theoretische Physik und Astrophysik, 
Christian-Albrechts-Universit\"at Kiel, D-24098 Kiel, Germany}
\author{N. Fitzer and R. Redmer}
\address{Fachbereich Physik, Universit\"at Rostock, 
Universit\"atsplatz 3, D-18051 Rostock, Germany}
\date{\today)\\(submitted to {\it Phys. Rev. B}}

\maketitle
\begin{abstract}
Quantum Monte Carlo methods have proven to predict atomic and bulk
properties of light and non-light elements with high accuracy. Here we
report on the first variational quantum Monte Carlo (VMC) calculations
for solid surfaces.  Taking the boundary condition for the simulation
from a finite layer geometry, the Hamiltonian, including a nonlocal
pseudopotential, is cast in a layer resolved form and evaluated with a
two-dimensional Ewald summation technique.  The exact cancellation of
all Jellium contributions to the Hamiltonian is ensured.  The many-body
trial wave function consists of a Slater determinant with parameterized
localized orbitals and a Jastrow factor with a common two-body term
plus a new confinement term representing further variational freedom to
take into account the existence of the surface.  We present results for
the ideal (110) surface of Galliumarsenide for different system sizes.
With the optimized trial wave function, we determine some properties
related to a solid surface to illustrate that VMC techniques provide
standard results under full inclusion of many-body effects at solid
surfaces.
\end{abstract}


\section{Introduction}
The theoretical description of solid surfaces bears fundamental aspects
as well as technologically important applications. Most of the
theoretical approaches are based on common local density functional
theory (DFT-LDA) or its improvements (GGA). Despite its success a
systematic development is hampered by the inherent approximations made,
especially for highly correlated or inhomogeneous systems. Therefore,
alternative approaches are desirable to test and further develop the
quality of the exchange-correlation functional.  It is a long tradition 
to use Quantum Monte Carlo (QMC) techniques for the homogeneous electron
gas\cite{CeAl80} not only to describe the system itself, but also to
apply the results as an approximation to inhomogeneous systems. 
Especially through the use of nonlocal pseudopotentials 
\cite{FWL88,FWL90B} within variational quantum Monte Carlo (VMC), 
the extension of QMC methods from homogeneous systems to solids of
non-light elements is state of the art with a variety of aspects such 
as cohesive properties of elemental \cite{WKRJNFFM96} or bipolar
\cite{EcSc96,MFB97} solids, or excitation energies.\cite{MiMa94,%
KHTNR98,KHWNFR99}

Here we propose a scheme to apply the VMC method to solid surfaces.
Being safely founded on Ritz's variational principle, all
approximations made in the trial wave function are controlled. An
unknown functional as in DFT or a perturbative approach like a
diagrammatic many-body expansion as in GW is thereby avoided.

The paper is organized as follows. In Sec.~\ref{method}, we describe
the method with a detailed discussion of the boundary conditions, the
Hamiltonian and the trial wave function. In Sec.~\ref{results}, we
present results for the optimization of the confinement term for two
different system sizes keeping other surface specific parameters in the
one-body orbitals fixed.  The dependence of the total energy on the
coordinates of a surface ion is examined to test the sensitivity of the
method to surface specific information.

\section{Method}
\label{method}

\subsection{Stochastic Integration}
In VMC the expectation value of the 
Hamiltonian of a quantum system in the state $\Psi_{\lambda}$ 
\begin{eqnarray}
\langle \hat H \rangle & = &
 \int \frac{{| \Psi_{\lambda}(\vec Y)|}^2}{%
             \int {| \Psi_{\lambda}(\vec Y)|}^2
              \, d\vec Y}
        \frac{\hat H(\vec Y)\Psi_{\lambda}(\vec Y)}{%
         \Psi_{\lambda}(\vec Y)} \, d\vec Y
\end{eqnarray}
is calculated by stochastic integration over the coordinates $\vec Y$
of the system, where $\vec \lambda $ denotes a set of parameters
representing the variational freedom in the trial wave function. In
practice the Metropolis algorithm~\cite{Met53} is used to guarantee
importance sampling during the random walk so that the energy can be
estimated safely as the mean of the local energy with purely
statistical error
\begin{equation}
\langle \hat H \rangle \approx E_{\lambda} := \frac1M \sum_{i=1}^{M} 
 \frac{\hat H(\vec Y_i)\Psi_{\lambda}(\vec Y_i)}{%
  \Psi_{\lambda}(\vec Y_i)}\, \mbox{,}
 \qquad
  \sigma(E_{\lambda})=
   \sqrt{%
    \frac{1}{M} 
      \mbox{\rm Var}(\frac{\hat H \Psi_{\lambda}}{\Psi_{\lambda}})
   } \; \mbox{.}
\end{equation}
Ritz's variational principle ensures that the minimum 
$E_{\lambda^\ast}$ of $E_{\lambda}$ in the parameter space is an upper 
bound for the true ground state energy $E_0$. 

\subsection{Boundary conditions for a solid surface}
\label{boundaryconditions}
The central problem in a QMC simulation of a solid surface is the
choice of the boundary conditions along the surface normal which is
chosen here to be the $z$-axis of the coordinate system. Often a
supercell geometry is used to model the semi-infinite crystal: the
simulation cell contains a slab of layers surrounded by vacuum on both
sides in the $z$-direction, and periodic boundary conditions are
applied to all three directions. The use of a localized basis eases the
choice of suitable boundary conditions. We assume periodic boundary
conditions only in the surface plane and take a symmetric slab of
layers with the normalizability of the wave function as the only
boundary condition in the $z$-direction. This layer resolved
formulation can be generalized to the semi-infinite crystal as well.
The Hamiltonian (see Sec.~\ref{hamiltonian}) and the wave function (see
Sec.~\ref{wavefunction}) are both affected by the choice of the boundary
conditions. First, we describe the adaption of the geo\-metry of the
simulation cell to the (110) surface of GaAs.

With the $z$-axis along the $[110]$-direction and the $x\mbox{,}y$-axes 
in the $(110)$-plane, the following set of vectors spans the lattice of 
GaAs: $ 
\vec a_1 = a\,(\frac{1}{\sqrt{2}},0,0)\mbox{,}\, 
\vec a_2 = a\, (0,1,0)\mbox{,}\, 
\vec a_3 = a\, (\frac{1}{\sqrt{8}},\frac12,\frac{1}{\sqrt{8}}) 
$, the basis consisting of $\vec \tau_{\rm As} = (0,0,0)$ and $\vec 
\tau_{\rm Ga} = a\,(\frac{1}{\sqrt{8}},\frac14,0)$, where $a$ is the 
lattice constant of cubic GaAs. The two-dimensional lattice 
$\{\vec R^{\nu}_{\|}\}$ of the simulation cell is chosen as the span of 
$ \vec b_1 = \mbox{\rm NX}\cdot \vec a_1\mbox{, } 
\vec b_2 = \mbox{\rm NY}\cdot \vec a_2\mbox{.} $
Denoting the number of lattice planes inside the simulation cell by
{\rm NZ}, we completely characterize the cell by the triple ({\rm NX,
NY, NZ}) containing $\mbox{\rm N}=8\cdot\mbox{\rm NX}\cdot \mbox{\rm
NY}\cdot \mbox{\rm NZ}$ electrons and $\mbox{\rm K}=2\cdot \mbox{\rm
NX}\cdot \mbox{\rm NY} \cdot \mbox{\rm NZ}$ ions. {\rm NZ} must be
large enough to prevent artificial interactions between the two
surfaces. Formally the size of the simulation cell in $z$ is infinite
because no further assumptions about the $z$-direction are made. That
is why no interaction between the two surfaces through the vacuum can
occur as in the usual supercell-slab geometry.

To model the infinite solid in this geometry, taking as the third
primitive vector $ \vec b_3 = \mbox{\rm NZ}\cdot (\vec
a_3)_{\bot}=\mbox{\rm NZ}\cdot a \cdot (0,0,\frac{1}{\sqrt{8}}) $ gives
an orthorhombic simulation cell $\Omega$ with the constraint {\rm NZ}
to be even.

\subsection{Hamiltonian}
\label{hamiltonian}
The Hamiltonian of the considered system, here electrons of a solid or
of a solid surface in the Born-Oppenheimer approximation, consists of
the kinetic energy operator $\hat T$ of the electrons and the potential
energy operator $\hat V$ of all the Coulomb interactions in the
system.  For solids of non-light elements the inner shell electrons and
the nucleus can be replaced by a pseudo-ion with effective charge $Z$
and a nonlocal electron-ion pseudopotential for the valence
electrons.\cite{FWL90B} The potential energy operator reads
\begin{equation}
\hat V = \hat V_C + \hat V_{PP}^{\rm l} + \hat V_{PP}^{\rm nl}\;\mbox{,}
\end{equation}
where $\hat V_C$ is the operator of the remaining Coulomb interactions 
and $\hat V_{PP}^{\rm l}$ and $\hat V_{PP}^{\rm nl}$ are the local 
resp. nonlocal part of the pseudopotential. The nonlocal part is 
evaluated in a semi-local form.\cite{EcSc96} The resulting Coulomb 
energy like operator collects together the constant ion-ion-energy, 
some part of the electron-ion interaction and the full 
electron-electron interaction
\begin{equation}
\hat V_C = \hat V_{ii} + \hat V_{ei} + \hat V_{ee}\; \mbox{,}
\end{equation}
thereby including the contributions of the $Z/r$-asymptotic of 
$\hat V_{PP}^{\rm l}$ among $\hat V_{ei}$. The remaining 
pseudopotential part is short ranged, and only $\hat V_C$ is affected
by the boundary conditions.  To account for the surface plane parallel
periodic boundary conditions, a super-lattice with the vectors $\{\vec
R^{\nu}_{\|}\}$ is introduced, which adds to an arbitrary position of
charge its periodic counterparts.  For clarity and later reference,
$\hat V_C$ explicitly reads
\begin{eqnarray}
\label{vcoulomb}
  2 \hat V_C & := & 
    \sum_i \left[
      \sum_{j\ne i} \left(
        \sum_{\nu} \frac{1}{| \vec r_i - \vec r_j - \vec R^{\nu}_{\|} |}
      \right)
      + \sum_{\nu\ne 0} \frac{1}{| \vec R^{\nu}_{\|} |}
    \right] 
    \\
    & & 
   -\sum_i 
      \sum_{\alpha} 
        \sum_{\nu} \frac {Z_\alpha}{| \vec r_i - \vec R_{\alpha} - 
                                      \vec R^{\nu}_{\|} |}
      \nonumber
    \\
    & &
   -\sum_{\alpha}  
      \sum_i 
        \sum_{\nu} \frac {Z_\alpha}{| \vec R_{\alpha} - \vec r_i - 
                                      \vec R^{\nu}_{\|} |}
      \nonumber
    \\
    & &
    +\sum_{\alpha} \left[
      \sum_{\alpha'\ne \alpha} \left(
        \sum_{\nu} \frac{Z_\alpha Z_{\alpha'}}{%
                         | \vec R_{\alpha} - \vec R_{\alpha'} - 
                           \vec R^{\nu}_{\|} |}
      \right)     
      + \sum_{\nu\ne 0} \frac{Z_\alpha^2}{| \vec R^{\nu}_{\|} |}
    \right]\, \mbox{,} 
    \nonumber
\end{eqnarray}
where $\alpha$ runs over the index set of gallium and arsenic ions and
$i$ over the electrons in the simulation cell.  As in the
three-dimensional case each of these two-dimensional lattice summations
over $\nu$ diverges for every $\vec r$.

One possibility to overcome this is to introduce a neutralizing
background charge in every interaction. To ensure that all these
jellium contributions do not artificially change the total Coulomb
energy, the background charges are forced to cancel out locally, at
every point, by adding the following terms to $\hat V_C$, which 
vanish due to charge neutrality
\begin{eqnarray}
\label{zerojelliumi}
  0 & = & 
        (N - \sum_{\alpha} Z_{\alpha} ) \times
        \sum_{i} \sum_{\nu} 
          \frac{1}{\Omega} 
          \int \limits_{\Omega} 
            \frac{-1}{%
                  | \vec r_i - \vec R^{\nu}_{\|} - \vec R |}
          \; d^3R
  \\
\label{zerojelliuma}
  & = &
        (-N + \sum_{\alpha'} Z_{\alpha'}) \times
        \sum_{\alpha} Z_{\alpha} \sum_{\nu} 
          \frac{1}{\Omega} 
          \int \limits_{\Omega} 
            \frac{-1}{%
                  | \vec R_{\alpha} - \vec R^{\nu}_{\|} - \vec R |}
          \; d^3R \, \mbox{,}
\end{eqnarray}
where  $\vec R=(\vec R_\|,R_z)$. Splitting up the r.~h.~s.\ of
eq.~(\ref{zerojelliumi}) into two terms, these are fed into the first
resp.\ second sum over $i$ of eq.~(\ref{vcoulomb}). In the same way the
parts of eq.~(\ref{zerojelliuma}) are distributed over the sums over
$\alpha$. Now all interactions of eq.~(\ref{vcoulomb}) converge and can
be calculated separately.  We define the following two-dimensional {\bf
surface lattice potential}~$v^{s}$
\begin{eqnarray}
v^{s}(\vec r) & := &
\sum_{\vec R^{\nu}_{\|}}
 \Bigl(
  \frac{1}{| \vec r - \vec R^{\nu}_{\|} |}
  + J^{\nu}(\vec r)
 \Bigr) - \frac{1}{%
            | \vec r|}
\end{eqnarray}
with
\begin{eqnarray}
J^{\nu}(\vec r) & := &
   \frac{1}{z_0\cdot\Omega_{0}^{s}} \! \!
   \int \limits_{-\frac{z_0}{2}}^{+\frac{z_0}{2}}
   \!\int \limits_{\Omega_{0}^{s}} 
    \frac{-1}{|\vec r-\vec R^{\nu}_{\|} -\vec R |}
   \; d^2 R_\| dR_z\mbox{,}
\end{eqnarray}
where $\Omega_0^s$ is the area of the surface unit cell, and $z_0$
chosen so that $\Omega=\Omega_0^s \times
[-\frac{z_0}{2},+\frac{z_0}{2}]$. By generalizing ideas of
Ref.~\onlinecite{Lu96} we evaluate $v^{s}$ by Ewald summation
technique in the form
\begin{equation}
\label{ewaldend}
\mbox{\parbox{0.9\textwidth}{$ \begin{array}{rcl}
\displaystyle
v^{s}(\vec r)
 & = & 
\displaystyle
  \sum_{\vec R^{\nu}_{\|}} \left(
     \displaystyle
     \frac{\textstyle 1}{\textstyle |\vec r - \vec R^{\nu}_{\|}|} -
     \displaystyle
     \frac{\textstyle 1}{ \textstyle
       |\vec r + \alpha\, 
       \mbox{sign}(z){\vec e}_z - \vec R^{\nu}_{\|}|} \right)
     - \frac{1}{|\vec r|} \nonumber\\*
  & + &
 \displaystyle
    \frac{2\pi}{\Omega_0^s}\displaystyle \sum_{\vec G_\|\not=0}
     \displaystyle
     \frac{\textstyle
      \exp\left(
       i\,\vec G_\|\cdot \vec r_\| -
      |\vec G_\||(|z|+\alpha)
      \right)
     }{\textstyle |\vec G_\||} \nonumber\\*[3mm]
  & - &
   \displaystyle
\frac{2\pi}{\Omega_0^s}
      \left\{ \begin{array}{l@{,\;}l}
             \alpha
              & z \not\in [-\frac{z_0}{2},+\frac{z_0}{2}] \\*
             \alpha +|z| - 
              \frac{1}{z_0}
              \left(
               \left(\frac{z_0}{2}\right)^2+z^2
              \right)
              & z \in [-\frac{z_0}{2},+\frac{z_0}{2}] \mbox{.}\\*
            \end{array}
      \right. 
\end{array}
$}}
\end{equation}

The result for $\hat V_C$ for the simulation of the surface system reads
\begin{eqnarray}
\label{vcoulges}
 \hat V_C & = & 
 \frac12
 \sum_{i,j} \left(
 v^{s}(\vec r_{ij}) \right) + \frac12 \sum_{i\not=j} 
 \frac{1}{|\vec r_{ij}|} \\
 & &{} - 
 \sum_{i,\alpha} \left( Z_{\alpha}
 v^{s}(\vec r_{i\alpha}) \right) -  \sum_{i,\alpha} 
 \frac{Z_{\alpha}}{|\vec r_{i\alpha}|} \nonumber \\
 & &{} + 
 \frac12
 \sum_{\alpha,\alpha'} \left( Z_{\alpha} Z_{\alpha'}
 v^{s}(\vec r_{\alpha\alpha'}) \right) + 
  \frac12 \sum_{\alpha\not=\alpha'} 
 \frac{Z_{\alpha} Z_{\alpha'}}{|\vec r_{\alpha\alpha'}|} \nonumber \\
& &{} + \hat V_J(\{z_i\},\{z_{\alpha}\})
 \, \mbox{,} \nonumber \\
 & = & \hat V_s + \hat V_J \, \mbox{,}
 \label{vcoulgesshort}
\end{eqnarray}
where $\vec r_{ab}=\vec r_a - \vec r_b$ and $z_{\alpha}=(\vec
R_{\alpha})_z$.  The Coulomb interaction is decomposed into a term
$\hat V_s$ depending on the inter-particle difference vector only,
which gives the main contribution to the total energy, and into a term
$\hat V_J$ which collects up all remaining terms. As is shown in
appendix \ref{app:vja}, these depend on $z$-coordinates only. Appendix
\ref{app:vjn} contains details of the numerical implementation.

The everywhere bounded, relatively smooth surface lattice potential
$v^{s}$ is used per tabulated interpolation with a Herman-Skilman mesh
in $z$-direction. The tabulation is necessary to make practical 
calculations feasible.

\subsection{Wave function}
\label{wavefunction}
Our ansatz for the wave function $\Psi = \Psi_{S} \times \Psi_{J}$ 
consists of a product $\Psi_S$ of Slater determinants and a Jastrow 
factor $\Psi_J$ with a two-body term and a confinement term
\begin{eqnarray}
\Psi(\vec r_1\sigma_1,\ldots, \vec r_N\sigma_N) & = &
 D^{\uparrow}(\vec r_1, \ldots, \vec r_{\frac{N}{2}})
 D^{\downarrow}(\vec r_{\frac{N}{2} +1}, \ldots, \vec r_N) \nonumber \\*
 \! \! \! \! \! \! \! \! & \times & \exp \Big(
     - \sum_{i<j} u(\vec r_{ij},\sigma_{ij}) 
     - \sum_{i} v_{cf}(\vec r_i)
     \Bigr)
\end{eqnarray}
with $\sigma_{ij}$ denoting parallel or anti-parallel spins. The 
two-body term $u$ is chosen in a periodic form 
\begin{equation}
\label{jasu}
u(\vec r,\sigma) = A\Bigl(\frac{1}{r}+v^{s}(\vec r)\Bigr)
 \left( 1- e^{-r/F(\sigma)}
 \right) \; \mbox{.}
\end{equation}
$A$ is a variational parameter and $F(\sigma)$ is, for given $A$, fixed
by the cusp condition~\cite{Ka57} to compensate for the Coulomb
singularity.

The original formulation of Fahy, Wang, and Louie~\cite{FWL88,FWL90B}
included a one-body function $\chi(\vec r_i)$ in order to correct for
the effect of the two-body term $u$ of smoothing the electron charge
density, and to restore the LDA charge density. This idea has been
extended successfully to enlarge the variational freedom in the Jastrow
factor keeping the one-body states in the Slater determinant
unchanged.\cite{WKRJNFFM96} But one drawback of this approach is that
the node surface of $\Psi$ is entirely determined by $\Psi_S$ so that
e.g. further diffusion quantum Monte Carlo (DMC) calculations are
restricted to the quality of the node surface of the underlying VMC
wave function which in fact depends on the quality of the one-body
orbitals.

Our approach allows for a direct optimization of the one-body orbitals
in the Slater determinant~\cite{EcSc96} viz.
\begin{eqnarray}
\label{bulkorb}
  \phi_{\alpha \mu}(\vec r) & = & 
     \phi_{\alpha \mu}^{\rm As}(\vec r)
   + { \beta} \phi_{\tilde \alpha \mu}^{\rm Ga}(\vec r)
  \,\mbox{,} \\*
  \phi_{\alpha \mu}^{\rm As}(\vec r) & = & 
   {\gamma^{\rm As}} 
    \phi_{s}^{\rm As}([\vec r-\vec R_{\alpha}^{\rm As}]/{%
                                       \zeta_{s}^{\rm As}})
   + \sum_{i} {\tau}_{\mu}^{i} 
      {\phi}_{p_{i}}^{\rm As}([\vec r-\vec R_{\alpha}^{\rm As}]/
       {\zeta_{p}^{\rm As}}) 
  \,\mbox{,} \nonumber \\*
  \phi_{\tilde \alpha \mu}^{\rm Ga}(\vec r) & = & 
     {\gamma^{\rm Ga}} 
     \phi_{s}^{\rm Ga}([\vec r-\vec R_{\tilde \alpha}^{\rm Ga}]/
      {\zeta_{s}^{\rm Ga}})
   - \sum_{i} {\tau}_{\mu}^{i} {\phi}_{p_{i}}^{\rm Ga}
        ([\vec r-\vec R_{\tilde \alpha}^{\rm Ga}]/{\zeta_{p}^{\rm Ga}})
 \nonumber
\end{eqnarray}
for a zincblende system as GaAs.  The index $\alpha$ here only runs
over the set of arsenic ions in the simulation cell, $\mu$ over the
four tetrahedral directions $\vec\tau_{\mu}$ originating from an
arsenic ion, and $i$ over the Cartesian coordinates $(x,y,z)$. The
index $\tilde \alpha$ denotes the index of the corresponding gallium
ion for each bond and is fixed for given $\alpha$ and $\mu$. These
localized orbitals are parameterized hybrid bonds with seven
variational parameters $\beta,\gamma, \zeta$ in the bulk case.
Especially the contraction parameters $\zeta$ allow for compensation of
the Jastrow factor. The unmodified orbitals $\phi_{.}^{\rm
Ga},\phi_{.}^{\rm As}$ are $4s$- and $4p$- pseudo wave functions
corresponding to the nonlocal ab-initio LDA, generalized
norm-conserving atomic pseudopotential following 
Hamann.\cite{HSC79,Ham89}

In the construction of the localized one-body orbitals for the arsenic
ions in the outermost layer the corresponding gallium bond partners are
missing and new parameterized surface states have to be set.

Our ansatz for the one-body surface states consists of a doubly
occupied arsenic dangling bond orbital. It formally arises from the
bulk orbitals of eq.~(\ref{bulkorb}) by setting $\beta=0$ if $\alpha$
denotes an arsenic ion in the outermost layer and $\mu=\mu_{db}$
denotes a tetrahedral direction towards the vacuum. Apart from
$\gamma^{\rm DB}_{p}, \zeta_{s}^{\rm DB}$, and $\zeta_{p}^{\rm DB}$ as
three new surface specific composition and contraction parameters we
allow for the variation of the direction of the dangling bond orbital
by introducing two parameters, the angles $\vartheta$ and $\phi$,
determining the prefactor ${\tau}_{\mu_{db}}^{i}$. To allow for effects
of relaxation, i.e. the deviation of the ions from their ideal
positions, in the construction of the hybrid orbitals the prefactors
${\tau}_{\mu}^{i}$ are chosen proportional to the vector connecting two
neighboring ions.

With this variational freedom one would expect that increasing the
number of layers the total energy per atom comes asymptotically closer
to the theoretical bulk value, because the surface to bulk fraction
decreases.  But, as will be stated in Sec.~\ref{sec:solidsurface},
the energy actually increases instead of decreasing with increasing
system size. To compensate for the observed overshoot in smoothing 
the electron density over the whole range of the simulation cell by the
Jastrow factor, we introduce the following confinement term in the
Jastrow factor
\begin{equation}
v_{cf}(\vec r)\equiv v_{cf}(z)= \frac12  \cdot V_{cf}
 \Bigl[\tanh(S_{cf}(|z|-z_{cf}- L_{cf}))+1\Bigr]
\mbox{,}
\label{vcf}
\end{equation}
where $z_{cf}$ is the $z$-distance of the outermost layer from the 
center.

\section{Results and Discussion}
\label{results}

\subsection{Bulk}
\label{sec:Bulkresults}
In order to test the numerical implementation of the layer resolved
formulation of the Hamiltonian, of the geometry of the simulation cell,
and of the construction of the trial wave function by comparison with
independently obtained results, a bulk GaAs system is modeled by an
orthorhombic cell called (2,2,8) containing $256$ electrons. Usual 
three-dimensional boundary conditions are applied for the random walk, 
but two-dimensional boundary conditions for the Coulomb potential and a 
direct summation over the copies in $z$-direction, which can be 
truncated after a small number of copies because the potential is 
vanishing rapidly with increasing $z$-distance. The total energy is
determined for different cubic lattice constants $a$ keeping all other
variational bulk parameters fixed. The resulting minimal energy per
unit cell $E^{\rm b}=(-232.92 \pm 0.05)$ eV/u.c. at an equilibrium
lattice constant of $a^{\rm b}_0=(10.69 \pm 0.05) \, \mbox{a.u.}$ is in
very good agreement with our former results of $E^{\rm ref}=(-233.04
\pm 0.08 )$ eV/u.c.  at an equilibrium lattice constant of $a^{\rm
ref}_0= (10.69 \pm 0.1) \, \mbox{a.u.}$ obtained within the usual VMC
scheme for solids for the same number of electrons.\cite{EcSc96}

We conclude from this comparison that, with respect to numerical
accuracy, the layer resolved implementation is suitable to tackle the
many-body problem for solid surfaces.

\subsection{Solid Surface}
\label{sec:solidsurface}
First, VMC calculations without confinement terms were performed. Only
technically necessary changes due to the new form of the simulation
cell, due to the prescribed two-dimensional periodicity concerning the
Hamiltonian, the trial wave function and the random walk, and due to
the lack of binding partners for ions at the outermost layers, were
made. One would expect that with increasing number of layers the energy
should converge to the asymptotic of the (theoretical) bulk energy
$E^{\rm b}$. But for the surface system (2,2,8) an energy of $E^{\rm
s}=(-154.13 \pm 0.16)$ eV/u.c. is obtained.  What is even worse,
increasing the number of layers up to $12$, the total energy increases
monotonically with an average slope of 24 eV/layer.  So a
straightforward, brute force usage of the traditional VMC scheme seems
not possible.

\begin{figure}[h!]
 \mbox{%
  \psfig{figure=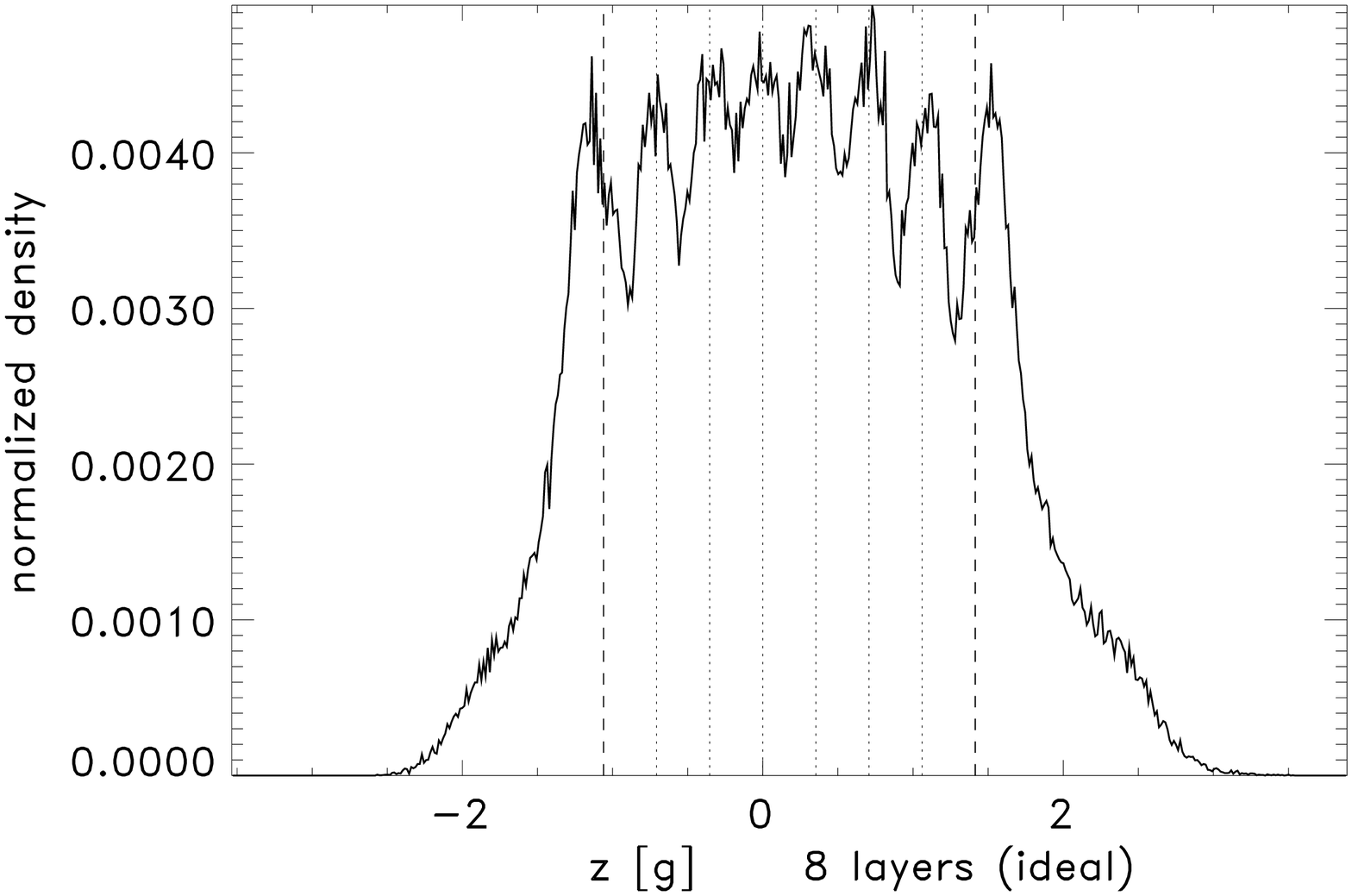,width=0.5\textwidth}
  \psfig{figure=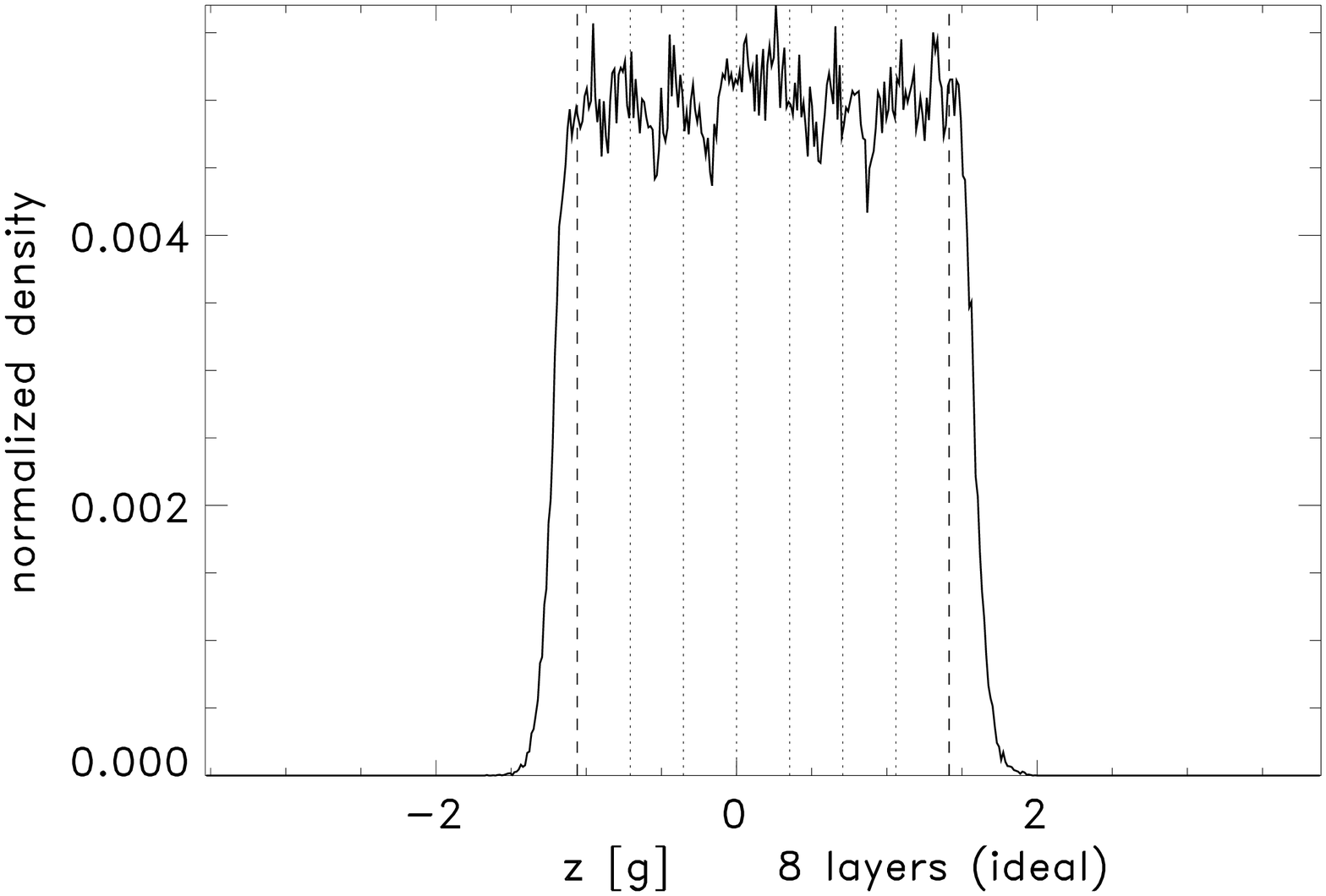,width=0.5\textwidth}
 }
  \caption{%
    $z$-resolved density with (right) and without (left) confinement 
    term in the trial wave function}
  \label{zsampling}
\end{figure}

Analyzing the different contributions to the total energy, one finds
that with respect to the bulk case, a per unit cell decrease in the
positive electron-electron interaction for larger systems is
drastically over-compensated by an increase in the negative
electron-ion interaction. This trend is even sharpened for bigger
system sizes.  Such drastic changes seem to be caused by electrons
drifting extraordinarily far into the vacuum region. The actual
existence of this mechanism is proved by observation of the
$z$-resolved density. As one can see from the left part of 
Fig.~(\ref{zsampling}) the electron density is smeared out into the 
vacuum regions. This behaviour comes from the known effect of the two-body
part of the Jastrow factor of smoothing the electron density. In the
bulk case, we have compensated for this effect through the introduction
of contraction parameters in the localized orbitals of the Slater
determinant.  But here, having allowed the outermost, arsenic dangling
bond orbitals to be contracted, we find no significant improvement on
the relevant scale of $10$~eV.

Taking advantage of the flexibility of VMC as a variational procedure,
we introduce an additional factor in the trial wave function allowing
for a compensation of the Jastrow pressure on the scale of the whole
surface system.  Formally, this factor belongs to the Jastrow term in
the form of a $z$-dependent confinement function $v_{cf}$.

To examine the behaviour of the total energy with respect to the
variation of the three confinement parameters, calculations for the
system (2,2,8) with $256$ electrons were performed yielding a minimal
energy of $E^{\rm s}_{256}= (-223.00 \pm 0.99)$ eV/u.c. for
($V_{cf},S_{cf},L_{cf})=(15.15 \pm 1.64, 0.137 \pm 0.013,3.01 \pm
0.77)$. This is an essential improvement in comparison to the former
results. The optimization was done by a weighted regression where the
total energy is written as a quadratic function of the variational
parameters.

In the region of parameter space where the global minimum is actually
located, the information of one of the three parameters is redundant
due to the presence of noise in form of the statistical uncertainty of
the Monte Carlo data $E_{\lambda}$. It is sufficient to fix one
parameter at an approximately correct value. E.g. setting $L_{cf}=3.0
\, \mbox{a.u.}$ and scanning the remaining two-dimensional parameter
space yields a fitted energy of $\tilde E^{\rm s}_{256}= (-222.92 \pm
2.16)$ eV/u.c. for $(V_{cf},S_{cf},L_{cf})=(15.27 \pm 0.24, 0.139 \pm
0.006,3.0)$. Finally, gathering statistics yields $\bar E^{\rm
s}_{256}=(-222.86 \pm 0.05)$ eV/u.c.  We conclude that the actual
uncertainty in energy is of the order of $100$ meV/u.c. and that the
trial wave function is essentially improved. Computed with the
optimized trial wave function, now the $z$-resolved density as seen in
the right part of Fig.~(\ref{zsampling}) is much more localized to
the solid proving that the above reasoning to explain the failure of
the bulk derived trial wave function was correct.

Whether the new confinement term is generally suitable also for larger
systems, the trend of increasing energy with number of layers should be
reversed.  Therefore we performed calculations for the system (2,2,12)
with $12$ layers and $384$ electrons. Again a two-dimensional scan of
the total energy landscape was performed yielding an fitted optimal
energy of $\tilde E^{\rm s}_{384}= (-223.64 \pm 2.09) $ eV/u.c. for
$(V_{cf},S_{cf},L_{cf})=(32.68 \pm 0.21, 0.103 \pm 0.001,3.0)$.
Repeated sampling with these values of parameters gives $\bar E^{\rm
s}_{384}=(-223.52 \pm 0.06) $ eV/u.c.

So by increasing the system size a slight decrease in total energy is
found which was originally to be expected from physical reasoning.
Comparing the optimal values for the variational parameters underlines
that the bigger systems needs a higher compensation of the Jastrow
pressure which is reflected in the corresponding values of $V_{cf}$ and
$S_{cf}$.

Further lowering in total energy, now on the scale of $1$ eV, has to be
achieved by other variational degrees of freedom, e.g. by variation of
parameters in the one-body orbitals in the Slater determinant. These
results will be presented in a forthcoming publication.

\begin{figure}[b!]
  \psfig{figure=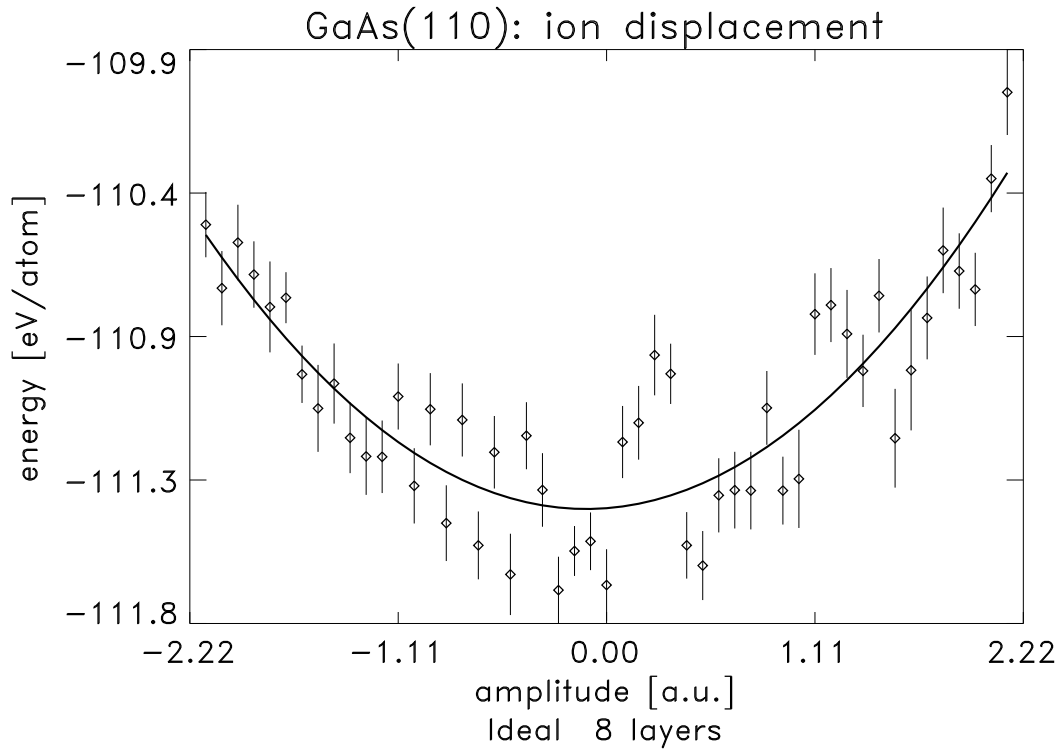,width=0.8\textwidth}
  \caption{%
    Total energy in [eV/atom] of slab of 8 layers with outermost
    arsenic ions displaced within layer plotted against displacement
    in atomic units. Error bars denote the statistical uncertainty 
    $\sigma(E_{\lambda})$ of individual Monte-Carlo energies 
    $E_{\lambda}$. Solid line shows the best fitting parabola.}
  \label{ionrelax}
\end{figure}

Here in the rest of this section we focus on the question, to which
extent the proposed VMC procedure can resolve surface specific
information. As a test case an in-layer displacement of the arsenic
ions in the outermost layer is chosen.  Is this statistically
resolvable within the current stage of optimization of the trial wave
function?  To improve the flexibility of the trial wave function with
respect to ion displacement, the coefficients of the $p_i$-orbitals are
chosen so that the resulting hybrid bond orbitals are automatically
concentrated in the bond direction keeping all other variational
surface specific parameters fixed.  In Fig.~(\ref{ionrelax}) the
total energy per unit cell is plotted against the displacement of an
arsenic ion perpendicular to the As-Ga chains of the ideal (110)
surface. Despite the presence of statistical noise a minimum is clearly
detectable. The equilibrium position is found very close to the ideal
position slightly moved to the direction away from the As-Ga chains. So
the determination of elements of the dynamical matrix via a parabolic
fit is statistically relevant within this VMC procedure.

In conclusion, we proposed a scheme to describe solid surfaces in the
frame of VMC techniques in order to use their fundamental aspects also
in this field.  Starting from the bulk derived, only slightly modified
trial wave function, showed a failure with respect to finite size
analysis. The reason was found in a large drift of the electron density
into the vacuum. By introducing a new confinement term to the trial
wave function much better results, with respect to the absolute value
in total energy and with respect to the size dependence, were obtained.
The sensitivity and relevance of the technique to surface structures
could be proved showing that a surface analysis is feasible. Further
systematic surface specific improvements in the trial wave function
should allow to start from a well founded, variational basis to finally
yield surface specific statements, as relaxation or reconstruction, and
many-body influences therein.

\section{Acknowledgments}
We thank D.~Lukas and D.~Schulz for valuable discussions.
Support from the Deutsche Forschungsgemeinschaft under 
Grants No.~Scha 360/8-1 and No.~Scha 360/17-1 is gratefully 
acknowledged.  The calculations were performed on the {\sc Cray-T3E} 
at the Konrad-\-Zuse-\-Zentrum f\"ur Informationstechnik, Berlin.

\appendix

\section{Explicit form of $V_J$}
\label{app:vja}
After insertion of eqs. (\ref{zerojelliumi}) and (\ref{zerojelliuma})
into eq. (\ref{vcoulomb}) and regrouping, four terms like

\begin{eqnarray}
\label{app:vcvj}
  2 V_{ab} 
  & = & 
    \sum_{\gamma}^{N_a}
      \sum_{\gamma'\ne\gamma}^{N_b}
        \sum_{\nu} 
        (
         \frac{1}{|\vec r_{\gamma} - \vec r_{\gamma'} - 
            \vec R^{\nu}_{\|}|}
          +  \frac{1}{\Omega} 
             \int \limits_{\Omega} 
               \frac{-1}{%
                    | \vec r_{\gamma} - \vec R^{\nu}_{\|} - \vec R |}
             \; d^3R
        ) \\
  & + & 
    \sum_{\gamma}^{N_a}
      [
        \sum_{\nu \ne 0} 
        (
         \frac{1}{|\vec R^{\nu}_{\|}|}
          +  \frac{1}{\Omega} 
             \int \limits_{\Omega} 
               \frac{-1}{%
                    | \vec r_{\gamma} - \vec R^{\nu}_{\|} - \vec R |}
             \; d^3R
        ) 
        + \int \limits_{\Omega} 
             \frac{-1}{%
                  | \vec r_{\gamma} - \vec R |}
             \; d^3R
      ] \nonumber \\
  & =: &
\label{app:vcvj0}
    \sum_{\gamma}
      \sum_{\gamma'}
        {\sum_{\nu}}'
        (
         \frac{1}{|\vec r_{\gamma} - \vec r_{\gamma'} - 
           \vec R^{\nu}_{\|}|}
          +  \frac{1}{\Omega}
             \int \limits_{\Omega}
               \frac{-1}{%
                    | \vec r_{\gamma} - \vec R^{\nu}_{\|} - \vec R |}
             \; d^3R
        ) \; \mbox{.}
\end{eqnarray}
occur, where $\gamma$ can either run through an electron index set or
an ion index set, read $\gamma=\alpha$ resp. $\gamma=i$ in
eq.~(\ref{vcoulomb}).  The prime at the summation over $\nu$ means that
the divergent term $1/|\vec R^{\nu=0}_{\|}|$ has to be  skipped in the
$\nu=0$ term of the sum if the outer indices $\gamma$ and $\gamma'$ are
equal and denote the same particle, however, the associated integral
still being kept.

Substituting $\vec R' := \vec R^{\nu}_{\|} + \vec R $ and skipping the
prime, the volume of integration changes from $\Omega = \Omega_0^s
\times [-\frac{z_0}{2}\,,\,+\frac{z_0}{2}]$ to $\Omega_{\nu} := \Omega
+ \vec R^{\nu}_{\|} =: \Omega^{s}_{\nu} \times
[-\frac{z_0}{2}\,,\,+\frac{z_0}{2}]$. Defining $\Omega_{\infty} :=
(\sum_{\nu} \Omega^{s}_{\nu}) \times
[-\frac{z_0}{2}\,,\,+\frac{z_0}{2}]$ eq.~(\ref{app:vcvj0}) now reads
\begin{eqnarray}
\label{app:vcvj1}
  2V_{ab}
  & = &
    \sum_{\gamma}
      \sum_{\gamma'}
      [
        {\sum_{\nu}}'
        (
         \frac{1}{|\vec r_{\gamma} - \vec r_{\gamma'} - 
           \vec R^{\nu}_{\|}|}
        )
          +  \frac{1}{\Omega}
             \int \limits_{\Omega_{\infty}}
               \frac{-1}{%
                    | \vec r_{\gamma} - \vec R |}
             \; d^3R
      ] \; \mbox{.} 
\end{eqnarray}
The next step is to substitute $\vec R = \vec R' + \vec r_{\gamma'}$ in
the integrals and skipping the prime again. Now the integration runs
over $\Omega_{\infty}^{\gamma'}:= (\sum_{\nu} \Omega^{s}_{\nu}) \times
 [-\frac{z_0}{2}-z_{\gamma'}\,,\,+\frac{z_0}{2}-z_{\gamma'}]$.
Furthermore, using the additivity of integration yields
\begin{eqnarray}
\label{app:vcvj2}
  2V_{ab}
  & = & 
    \sum_{\gamma}
      \sum_{\gamma'}
      [
        {\sum_{\nu}}'
        (
         \frac{1}{|\vec r_{\gamma} - \vec r_{\gamma'} - 
           \vec R^{\nu}_{\|}|}
        )
          +  \frac{1}{\Omega}
             \int \limits_{\Omega_{\infty}}
               \frac{-1}{%
                    | \vec r_{\gamma} - \vec r_{\gamma'} - \vec R |}
            \; d^3R
      ]
      \\
  & & 
    +\sum_{\gamma}
      \sum_{\gamma'}
      [
    \frac{1}{\Omega} 
             \int \limits_{\Omega_{\infty}^{\gamma'} - \Omega_{\infty}}
               \frac{-1}{%
                    | \vec r_{\gamma} - \vec r_{\gamma'} - \vec R |}
             \; d^3R
      ] \; \mbox{.} \nonumber
\end{eqnarray}
Integration over $\Omega_{\infty}^{\gamma'} - \Omega_{\infty}$ means
that contributions from $\Omega_{\infty}^{\gamma'}$ count positively
while $\Omega_{\infty}$ contributes negatively. The first line of
eq.~(\ref{app:vcvj2}) depends only on the difference vector $\vec
r_{\gamma\gamma'} := \vec r_{\gamma} - \vec r_{\gamma'}$.  So it is
useful to define
\begin{eqnarray}
\label{app:vcvsdef}
  v^s(\vec r_{\gamma\gamma'})
  & :=  &
        \sum_{\nu}
        (
         \frac{1}{|\vec r_{\gamma} - \vec r_{\gamma'} - 
            \vec R^{\nu}_{\|}|}
          +  \frac{1}{\Omega}
             \int \limits_{\Omega}
               \frac{-1}{%
                    | \vec r_{\gamma} - \vec r_{\gamma'} 
                     - \vec R^{\nu}_{\|} - \vec R |}
            \; d^3R 
        )
        - \frac{1}{| \vec r_{\gamma\gamma'} | }
\end{eqnarray}
and
\begin{eqnarray}
\label{app:vcvjdef}
  \varphi^J(z_{\gamma},z_{\gamma'}) 
  & := &
    \frac{1}{\Omega}
    \int \limits_{\Omega_{\infty}^{\gamma'} - \Omega_{\infty}} 
      \frac{-1}{%
        | \vec r_{\gamma} - \vec r_{\gamma'} - \vec R |}
    \; d^3R
    \; \mbox{.}
\end{eqnarray}
Though $\vec r_{\gamma}$ and $\vec r_{\gamma'}$ formally appear on the
r.~h.~s.  of eq.~(\ref{app:vcvjdef}), the actual dependence on $\vec
r_{\|}$ drops out due to the infinite integration parallel to the
surface.  Changing the variable of integration to $\vec R'=\vec R +
\vec r_{\gamma'}$ in eq.~(\ref{app:vcvjdef}) leads to the result
\begin{equation}
\label{app:vcvj3}
  \varphi^J(z_{\gamma},z_{\gamma'})
  = 
    \frac{1}{\Omega}
    \int \limits_{\Omega_{\infty} - \tilde \Omega_{\infty}^{\gamma'}}
      \frac{-1}{%
        | \vec r_{\gamma} - \vec R |}
    \; d^3R
    \; \mbox{,}
\end{equation}
with $ \tilde \Omega_{\infty}^{\gamma'}:= (\sum_{\nu} \Omega^{s}_{\nu})
\times [-\frac{z_0}{2}+z_{\gamma'}\,,\,+\frac{z_0}{2}+z_{\gamma'}]$.
Physically this corresponds to a dipole like charge configuration with
films of finite width. A compact solution of this problem of
electrostatics is given in appendix~\ref{app:vjn}.  Finally, inserting
eqs.~(\ref{app:vcvsdef}) and (\ref{app:vcvj3}) in
eq.~(\ref{app:vcvj2}), with additional prefactors
$Z_{\gamma},Z_{\gamma'}$ to account for arbitrarily charged particles
in the general case, $2V_{ab}$ can be expressed as
\begin{eqnarray}
\label{app:vcvjfinal}
  2V_{ab}
  & = & 
   (
    \sum_{\gamma}
    \sum_{\gamma'} Z_{\gamma}Z_{\gamma'}
      v^s(\vec r_{\gamma\gamma'}) + 
    \sum_{\gamma}
    \sum_{\gamma'\neq \gamma} Z_{\gamma}Z_{\gamma'}
    \frac{1}{| \vec r_{\gamma\gamma'} |}
  ) +
    \sum_{\gamma}
    \sum_{\gamma'} Z_{\gamma}Z_{\gamma'}
      \varphi^J(z_{\gamma},z_{\gamma'}) \\ 
\label{app:vcvjfinal2}
  & =: &
    2V_{ab}^s + 2V_{ab}^J
\; \mbox{.}
\end{eqnarray}
With these abbreviations the total Coulomb energy of 
eq.~(\ref{vcoulomb}) can be written as
\begin{eqnarray}
\label{app:vcdecomp}
  V_C
  & = & 
    V_{ee}^s + V_{ee}^J +
    V_{ei}^s + V_{ei}^J +
    V_{ie}^s + V_{ie}^J +
    V_{ii}^s + V_{ii}^J \\
  & = &
    V_{ee}^s +
    2V_{ei}^s +
    V_{ii}^s +
    V_{ee}^J +
    V_{ei}^J +
    V_{ie}^J +
    V_{ii}^J \\
  & =:& V_s + V_J \, \mbox{.}
\end{eqnarray}

\section{Numerical evaluation of $V_J$}
\label{app:vjn}
As defined in eq.~(\ref{app:vcvjfinal2}) $2V_J$ consists of terms like 
\begin{equation}
\label{app:vjab}
  2 V_{ab}^J = \sum_{\gamma_a \gamma_b} Z_{\gamma_a} Z_{\gamma_b} 
               \varphi^J(z_{\gamma_a},z_{\gamma_b})\,\mbox{.}
\end{equation}
Due to the symmetry of the problem $\varphi^J$ can be expressed by
\begin{equation}
\label{app:phij}
  \varphi^J(z_{\gamma_a},z_{\gamma_b})=
    \left\{ 
      \begin{array}{l@{,\;}l}
        \varphi^J_+(z_{\gamma_a};z_{\gamma_b})
          &  z_{\gamma_b} \geq 0 \\*
        \varphi^J_+(z_{\gamma_a}-z_{\gamma_b};z_{\gamma_b})
          &  z_{\gamma_b} < 0 \,\mbox{,}\\* 
      \end{array}
    \right.
\end{equation}
Before giving the explicit expression for $\varphi^J_+$ we define 
\begin{eqnarray}
  d_\rho(|z'|) & = & \mbox{min}\,(|z'|,z_0)\;\mbox{,} \\
  d_0(|z'|)    & = & \left|\, |z'| - z_0 \,\right| \;\mbox{.}
\end{eqnarray}

Then the potential generated by a dipole like charge distribution of
two oppositely charged films each of thickness $d_\rho$ with their 
inner borders separated by $d_0$ is given by

\begin{equation}
\label{app:phijp}
  \varphi^J_+(z;z') = \varphi_0 + 
    \frac{\mbox{sign}(z')}{\Omega_0^s z_0}\cdot
    \left\{
      \begin{array}{l@{\;}l}
      0
      & ,z \in (-\infty\,,\,-\frac{z_0}{2}] \\*
      \frac12 (z+\frac{z_0}{2})^2
      & ,z \in (-\frac{z_0}{2}\,,\,-\frac{z_0}{2}+d_\rho] \\*
      d_\rho (z+\frac{z_0}{2} - \frac{d_\rho}{2})
      & ,z \in (-\frac{z_0}{2}+d_\rho\,,\,-\frac{z_0}{2}+d_\rho+d_0] \\*
              -\frac{z^2}{2} + 
              (-\frac{z_0}{2}+2d_\rho+d_0)(z+\frac{z_0}{2}) + \\*
              \frac12 (\frac{z_0}{2})^2 - d_\rho(d_\rho + d_0) - 
                \frac{d_0^2}{2}
      & ,z \in (-\frac{z_0}{2}+d_\rho+d_0\,,
                \,-\frac{z_0}{2}+2d_\rho+d_0]\\*
      d_\rho d_0 + d_\rho^2
      & ,z \in (-\frac{z_0}{2}+2d_\rho+d_0\,,\,+\infty) \, \mbox{.}\\*
     \end{array}
    \right.
\end{equation}
The potential is determined up to a constant $\varphi_0$. This is 
fixed by the convention that a charge at $z=-\infty$ feels zero 
potential leading to $\varphi_0=0$.

\end{document}